\documentclass{ws-ijmpa}

\begin{document}

\markboth{Authors' Names}
{Instructions for Typing Manuscripts (Paper's Title)}

%
\catchline{}{}{}{}{}
%

\title{STUDY OF THE THRESHOLD BEHAVIOR OF THE $\eta$N SCATTERING AMPLITUDE THROUGH THE ASSOCIATED PHOTOPRODUCTION OF
$\phi$- AND $\eta$-MESONS
}

\author{MADELEINE SOYEUR
}

\address{DAPNIA/SPhN, CEA/Saclay\\ F-91191 Gif-sur-Yvette Cedex, France\\
madeleine.soyeur@cea.fr}

\author{MATTHIAS F. M. LUTZ}

\address{GSI, Planckstrasse 1\\
D-64291 Darmstadt, Germany\\
m.lutz@gsi.de}

\maketitle

\begin{history}
\received{Day Month Year}
\revised{Day Month Year}
\end{history}

\begin{abstract}
We suggest that the $\gamma p \rightarrow \phi \eta p$
reaction cross section, in the kinematics where the $\eta p$ invariant mass in the final state
lies between the threshold value (m$_p$+m$_\eta$) and the N*(1535) resonance mass,
is largely determined by the $\eta N$ scattering amplitude close to threshold. The initial photon
energy is chosen in the range $4\, <\,E_\gamma^{Lab}\,<\,5$ GeV,
in order to reach low (absolute) values of the squared 4-momentum transfer from the initial photon
to the final $\phi$-meson. In these conditions, we expect the t-channel $\pi^0$-
and $\eta$-meson exchanges to drive the dynamics underlying the $\gamma p \rightarrow \phi \eta p$
process. We show that the $\eta$-exchange is the dominating contribution to the cross section
while the $\pi^0$-exchange is negligible.
The $\eta$-$\pi^0$ interference is of the order
of $20\,-\,30 \%$. The sign of this term is not known and alters significantly
our results. Data on the
$\gamma p \rightarrow \phi \eta p$ process would be therefore very useful to help unravelling
the behavior of the $\eta p$ scattering amplitude close to threshold and assessing the
possibility of producing $\eta$-nucleus bound states.
\end{abstract}

\ccode{PACS numbers: 13.30.Eg;13.60.Le;13.60.Rj;14.20.Gk}

\section{Introduction: the $\eta$N scattering amplitude and scattering length}

The main quantity of interest in this work is the s-wave $\eta$-nucleon scattering
amplitude (f$_{\eta N}$) displayed in Fig 1. This particular description of the real and imaginary
parts of the $\eta$-nucleon scattering amplitude is obtained in the re\-lativistic coupled-channel
scheme of Ref.~\refcite{Lutz1}.
Its main features are however
derived in other approaches as well, for example the K-matrix coupled-channel calculation
of Ref.~\refcite{GreenWycech}.

\begin{figure}[t]
\centerline{\psfig{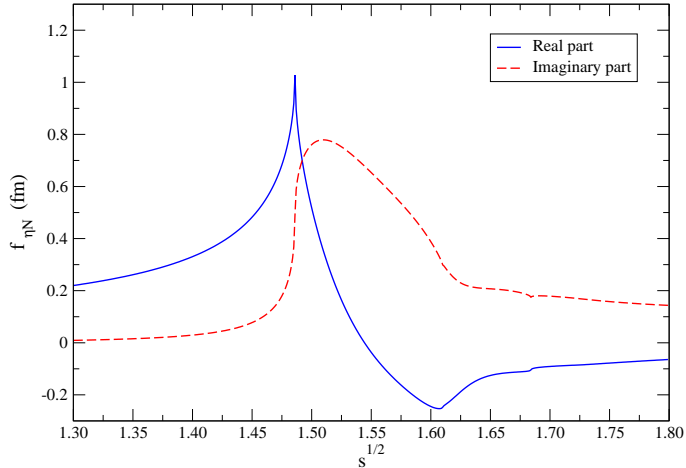}}
\vspace*{8pt}
\caption{Real and imaginary parts of the s-wave $\eta N$ scattering amplitude (from Ref. 1).}
 \label{f1}
\end{figure}

The most striking feature is the shape of the real part of the s-wave $\eta$-nucleon scattering
amplitude close to threshold ($\sqrt s$ = 1.486 GeV). The narrow 'spike' in the vicinity of the $\eta$-nucleon
threshold implies that physics will change very rapidly with the invariant $\eta N$ pair mass
in this region. In particular, the threshold value of Re f$_{\eta N}$ will decrease significantly in nuclei
where the $\eta$-meson scatters from bound nucleons. The ima\-ginary part of the amplitude is
dominated by the N*(1535) resonance and hence peaks close to the resonance nominal mass. Because
the N*(1535) resonance lies only 49 MeV above threshold, the invariant mass region sensitive
to the real part of the amplitude is quite restricted and limited to $\eta$-nucleon pair
total center of mass energy within $\sim$25 MeV of threshold. The $\eta$-nucleon scattering
amplitude displayed in Fig. 1 describes only s-wave dynamics and is therefore not valid beyond
$\sqrt s\approx$1.6 GeV~\cite{Lutz1}. For future considerations, it is useful to recall that
the N*$_{1/2^-}$(1535) has the same spin and isospin as the nucleon but the opposite parity.
The width of the N*(1535) is of the order of 150 MeV. Its two main decay channels
are the N$\pi$ (35-55~$\%$) and the N$\eta$ (30-55 $\%$) final states \cite{PDG}.

The $\eta$-nucleon scattering length in the model
of Ref.~\refcite{Lutz1} can be read off Fig.~1. It is $a_{\eta N}$=(1.03 + i 0.49) fm.
The corresponding quantity obtained in Ref.~\refcite{GreenWycech} is $a_{\eta N}$=(0.91(6) + i 0.27(2)) fm.
The range of values found in the literature for Re $a_{\eta N}$ (from $\sim$ 0.3 to $\sim$1 fm)
is however much broader than these two values suggest\cite{GreenWycech} and reflects the indirect nature of the determination of
 Re $a_{\eta N}$ from $\eta$-meson production
reactions such as $\pi N \rightarrow \eta N$, $\gamma N \rightarrow \eta N$, $p n \rightarrow \eta d$,
$p d \rightarrow \eta ^3He$, etc...

The purpose of this work is to show that the
$\gamma p \rightarrow \phi \eta p$ process in specific kinematic conditions
is most sensitive to the $\eta$-nucleon scattering
at threshold. These conditions are twofold. On the one hand
the $\eta p$ invariant mass in the final state
should be close to its threshold value (m$_p$+m$_\eta$) and not much larger than the N*(1535) resonance mass.
On the other hand the initial photon
energy should be sufficiently high, typically in the range $4\, <\,E_\gamma^{Lab}\,<\,5$ GeV,
in order to reach low (absolute) values of the squared 4-momentum transfer from the initial photon
to the final $\phi$-meson ($|t|<$ 1 GeV$^2$). The latter constraint is needed to ensure that
the description of the $\gamma p \rightarrow \phi \eta p$ reaction by a meson-exchange picture
be legitimate.

We discuss our t-channel meson-exchange picture of the $\gamma p \rightarrow \phi \eta p$ reaction
in Section 2. We present numerical results and concluding remarks in Section 3. A more extended
version of this study can be found in Ref.~\refcite{Lutz2}.

\section{The $\gamma \, p \rightarrow \phi \, \eta \, p$ reaction cross section
in the N*(1535) region}

In the restricted kinematics defined above, we expect t-channel meson-exchanges to drive
the dynamics underlying the $\gamma p \rightarrow \phi \eta p$
process. There are both experimental data and theoretical arguments to justify this
statement.

Recent measurements of the $\gamma \, p \rightarrow \phi \, p$ reaction
by SAPHIR at ELSA\cite{Barth} ($E_\gamma^{Lab}\,\leq$ 2.4 GeV), by CLAS at JLab\cite{CLAS}
($E_\gamma^{Lab}\,=$ 3.5 GeV) and LEPS at SPring-8\cite{Mibe} ($E_\gamma^{Lab}\,\leq$ 2.37 GeV)
all indicate that the differential cross sections $d\sigma/dt$ are peaked at forward angles.
This feature is more pronounced with increasing photon energy.
This behavior suggests that t-channel exchanges dominate the $\phi$ photoproduction dynamics at low t
for photon energies of a few GeV. The theoretical issue is to single out
the exchanges which are the most significant. In the case of the $\gamma \, p \rightarrow \phi \, p$ reaction,
scalar and pseudoscalar mesons as well as pomerons can be exchanged\cite{Titov}. The calculated cross sections,
and hence the relative importance of these exchanges, depend very much on the description of the vertices
(coupling constants and form factors). It is therefore quite difficult to disentangle the different
contributions.

In the case of the $\gamma p \rightarrow \phi \eta p$ process, the negative parity of the N*(1535)
implies that the photoproduction of a $\phi$-meson associated with the excitation
of the target proton into a N*(1535) resonance be induced by a pseudoscalar exchange. There are
only two terms, the $\pi^0$- and the $\eta$-exchanges depicted in Fig. 2.

\begin{figure}[h]
\begin{center}
$\begin{array}{c@{\hspace{1in}}c}
\multicolumn{1}{l}{\mbox{\bf }} &
    \multicolumn{1}{l}{\mbox{\bf }} \\ [-0.53cm]
\epsfxsize=2in
\epsffile{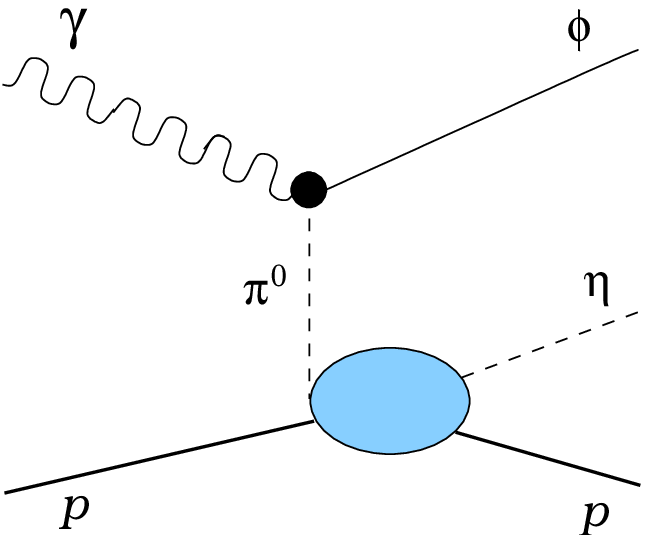} &
    \epsfxsize=2in
    \epsffile{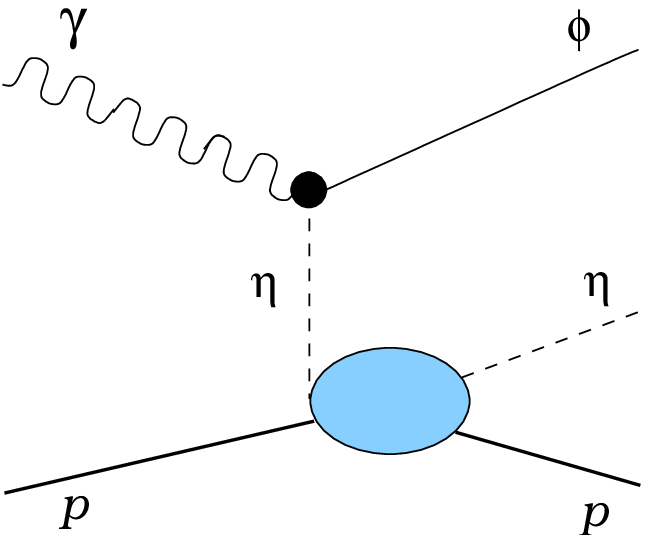} \\ [0.4cm]
\end{array}$
\end{center}
\caption{$\pi^0$- and $\eta$-exchange contribution to the $\gamma \, p \rightarrow \phi \, \eta \, p$ process.}
\label{f2}
\end{figure}

The dynamical inputs of our calculation are therefore the $\gamma\phi\pi^0$ and the $\gamma\phi\eta$ vertices
and the $\pi \, p \rightarrow \eta \, p$ and $\eta \, p \rightarrow \eta \, p$ scattering
amplitudes.

The $\gamma\phi\pi^0$ and the $\gamma\phi\eta$ vertices are described by the anomalous interaction Lagrangian
\begin{eqnarray}
{\cal L}^{int}_{\phi \chi \gamma} =
e \,\frac{g_{\phi \chi \gamma}} {2 m_\phi} \, \varepsilon^{
\mu\nu\alpha\beta} \, \phi_\mu \, (\partial_\nu \chi)\, F_{\alpha \beta},
\label{anomalousLintpi}
\end{eqnarray}
where $F_{\alpha \beta}$ is the electromagnetic field tensor,
\begin{eqnarray}
F_{\alpha \beta}\, =\partial_\alpha A_\beta \,-\,\partial_\beta A_\alpha,
\label{defF}
\end{eqnarray}
\noindent
and $\chi$ denotes the chiral pseudoscalar mesons of interest, the pion or the $\eta$-meson.
Using this interaction Lagrangian to calculate the $\phi\rightarrow \pi^0\gamma$
and $\phi\rightarrow \eta\gamma$ partial widths, we can derive the values of the
coupling constants from the measured partial widths\cite{PDG}. We find
\begin{eqnarray}
|g_{\phi\pi\gamma}|\, \simeq\, 0.13,\mkern 30 mu |g_{\phi\eta\gamma}|\, \simeq\, 0.70.
\label{coupling constants}
\end{eqnarray}
On the basis of these couplings only, the pion-exchange contribution to the cross section will be
suppressed by a factor of the order of 29 compared to the $\eta$-exchange contribution.
This suppression factor is about 5 for the $\pi^0-\eta$ interference. This interference may therefore
be significant. Its sign cannot be determined unambiguously from available data
(see the discussion of Ref.~\refcite{Lutz2}). We consider therefore both constructive and destructive
interferences in the presentation of the numerical results.

The
$\pi \, p \rightarrow \eta \, p$ and $\eta \, p \rightarrow \eta \, p$ s-wave scattering
amplitudes used to calculate the $\gamma p \rightarrow \phi \eta p$
cross section were obtained in the relativistic, unitary coupled-channel model of Ref.~\refcite{Lutz1}.
This approach reproduces a large set of pion-nucleon and photon-nucleon
scatte\-ring data in the energy range $1.4<\sqrt s <1.8$ GeV, in particular
the pion- and photon-induced $\eta$-meson production cross sections close to threshold.
The N*(1535) resonance of interest in this work is generated dynamically through
meson-baryon scattering using the Bethe-Salpeter equation.
The $\eta \, p \rightarrow \eta \, p$ amplitude is shown in Fig. 1
and the $\pi \, p \rightarrow \eta \, p$ amplitude in Fig. 3. One should note that
the s-wave $\pi \, p \rightarrow \eta \, p$ amplitude is significantly
smaller than the $\eta \, p \rightarrow \eta \, p$ amplitude close to the
$\eta$-meson threshold, suppressing further the pion-exchange contribution to the
$\gamma p \rightarrow \phi \eta p$ cross section.

The squared amplitudes corresponding to the $\pi$-exchange,
the $\eta$-exchange and their interference are
\begin{eqnarray}
\sum_{\lambda_\gamma, \lambda ,\bar
\lambda_\phi, \bar \lambda }\,\frac{1}{4}\,\mid M_{\gamma \,p\rightarrow \phi \, \eta \,p}^{\pi-exchange}\mid^2
&=&\frac {e^2\,g_{\phi\pi\gamma}^2} {4\,m_{\phi}^2}\frac{(m_\phi^2-t)^2}{(t-m_\pi^2)^2}\,\frac{1}{2}\,
\sum_{\lambda ,\bar
\lambda }\,|M_{\pi\,p \to \eta\,p}|^2,
\label{piexchange}
\end{eqnarray}
\begin{eqnarray}
\sum_{\lambda_\gamma, \lambda ,\bar
\lambda_\phi, \bar \lambda }\,\frac{1}{4}\,\mid M_{\gamma \,p\rightarrow \phi \, \eta \,p}^{\eta-exchange}\mid^2
&=&\frac {e^2\,g_{\phi\eta\gamma}^2} {4\,m_{\phi}^2}\frac{(m_\phi^2-t)^2}{(t-m_\eta^2)^2}\,\frac{1}{2}\,
\sum_{\lambda ,\bar
\lambda }\,|M_{\eta\,p \to \eta\,p}|^2,
\label{etaexchange}
\end{eqnarray}
\begin{eqnarray}
\sum_{\lambda_\gamma, \lambda ,\bar
\lambda_\phi, \bar \lambda }\,\frac{1}{4}\,\mid M_{\gamma \,p\rightarrow \phi \, \eta \,p}^{interference}\mid^2
=\frac {e^2\,g_{\phi\pi\gamma}\,g_{\phi\eta\gamma}} {4\,m_{\phi}^2}\frac{(m_\phi^2-t)^2}{(t-m_\pi^2)(t-m_\eta^2)}
\nonumber
\\\frac{1}{2}\,
\sum_{\lambda ,\bar
\lambda }\,(M^+_{\pi\,p \to \eta\,p}M_{\eta\,p \to \eta\,p}\,+\,M^+_{\eta\,p \to \eta\,p}M_{\pi\,p \to \eta\,p}).
\label{interference}
\end{eqnarray}

\begin{figure}[h]
\centerline{\psfig{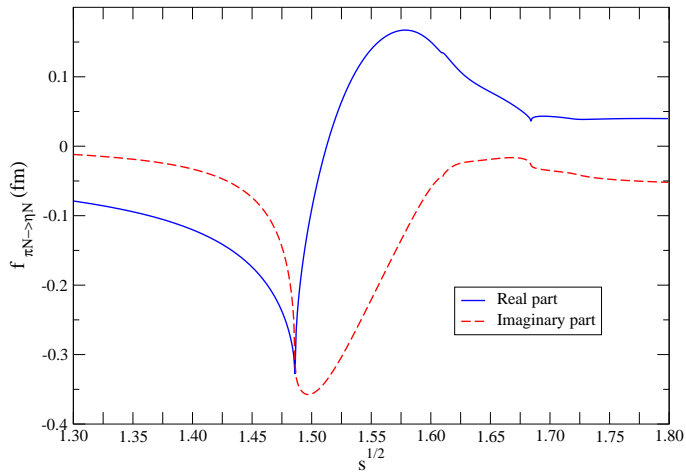}}
\vspace*{8pt}
\caption{Real and imaginary parts of the s-wave $\pi N \rightarrow \eta$N scattering amplitude (from Ref. 1).}
\label{f3}
\end{figure}

We assume in the derivation of these equations that the $\gamma \eta \phi$  and $\gamma \pi \phi$ transition form factors are one.
This is a choice based on dynamical considerations outlined in Ref.~\refcite{Lutz2} in the absence
of direct constraints from data.
There is an obvious uncertainty linked to that assumption.
If the $\gamma \eta \phi$  and $\gamma \pi \phi$ transition form factors turned out to be soft,
the differential cross sections, d$\sigma$/dt, could be significantly different.

Our model takes into account the $\eta$-nucleon final state interaction to all orders,
but does not treat $\phi$-nucleon rescattering in the outgoing channel.
We do not expect this rescattering to be very important in the kinematics under
consideration, i.e. with a large relative momentum between the
$\phi$-meson emitted at small angles and the recoiling target products.

Before showing results, it is useful to comment on the kinematics of the
$\gamma \, p \rightarrow \phi \, \eta \, p$ reaction.
At the $\phi$ threshold, $E_\gamma^{Lab}$=3 GeV and $|t_{min}|$ is 1.2 GeV$^2$.
This number is obtained assuming the
invariant mass of the $\eta$p pair to be the N*(1535) mass.
Meson-exchange models are expected to be valid at low $|t|$
($|t|<$ 1 GeV$^2$). We
consider therefore values of $E_\gamma^{Lab}$ significantly
above threshold. In this paper, we display results for
$E_\gamma^{Lab}$=4 GeV where $|t_{min}|$ is 0.38 GeV$^2$.
We refer to Ref.~\refcite{Lutz2} for a similar study at  5 GeV,
where $|t_{min}|$ is 0.26 GeV$^2$.

\section{Numerical results and concluding remarks}

We have calculated differential cross sections for the
$\gamma \, p \rightarrow \phi \, \eta \, p$ reaction at two values of
invariant mass of the $\eta p$ pair (denoted $\bar w$), chosen to be
sensitive to the real and to the imaginary parts of the s-wave $\eta N$ scattering amplitude.
We consider $\bar w$= 1.49 GeV (where the real and the imaginary parts are of comparable importance)
and  $\bar w$= 1.54 GeV (where the imaginary part is largely dominant).

We show first in Fig. 4 the differential cross section for the
$\gamma \, p \rightarrow \phi \, \eta \, p$ reaction
at $E_\gamma^{Lab}=\,$ 4 GeV  for a total center of mass energy of the $\eta p$ pair of 1.49 GeV.

\begin{figure}[h]
\centerline{\psfig{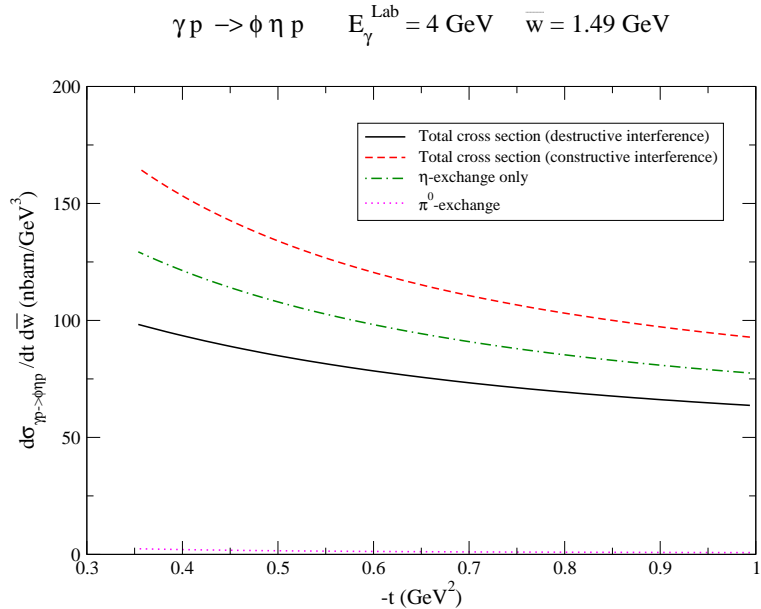}}
\vspace*{8pt}
\caption{Differential cross section
${d\sigma_{\gamma \,p\rightarrow \phi \, \eta \,p}}/ {dt \,d{\bar w}}$
at $E_\gamma^{Lab}=\,$ 4 GeV  for a total center of mass energy of the $\eta p$ pair of 1.49 GeV.
The full and dashed lines are the total differential cross sections assuming a destructive and a constructive
$\pi^0$-$\eta$ interference respectively. The dot-dashed and dotted lines show the contributions of the
$\eta$-meson and $\pi^0$-exchanges (from Ref. 4).}
\label{f4}
\end{figure}

The different curves illustrate the features and uncertainties of our calculation.
The full and dashed lines are the total differential cross sections assuming a destructive and a constructive
$\pi^0$-$\eta$ interference respectively.
The dot-dashed and dotted lines are the contributions of the
$\eta$-meson and of the $\pi^0$-exchanges. We remark first that $\pi^0$-exchange contribution
is negligible as expected. The $\eta$-meson exchange is clearly the dominant contribution to the
$\gamma \, p \rightarrow \phi \, \eta \, p$ cross section, hence justifying our primary motivation
to use that particular reaction to study the $\eta \, N \rightarrow \eta \, N$ amplitude
close to threshold. The full and dashed lines are indicative of the importance of the
$\eta-\pi^0$ interference. Depending on the relative sign of the coupling constants
$g_{\phi\pi\gamma}$ and $g_{\phi\eta\gamma}$, the differential cross section decreases
(like signs) or increases (opposite signs) by $\sim 20-30\%$.
If the $\eta-\pi^0$ interference is constructive, the pole structure of the $\eta$-exchange
and of the interference
leads to a rather sharp t-dependence close to $t_{min}$. This effect will increase with increasing
laboratory photon energy as a lower $|t_{min}|$ can be reached. If the $\eta-\pi^0$ interference is
destructive, the terms driving the increase of the differential cross section
at low $|t|$ cancel significantly, leading to a rather flat behavior.

In general, the cross section is rather small as a consequence of the value of $\bar w$,
which is very close to the $\eta N$ threshold. We refrain from performing an integration over $t$
in view of the fact that our model is not valid for $|t|>$ 1 GeV$^2$.

We present in Fig. 5 the differential cross section for the
$\gamma \, p \rightarrow \phi \, \eta \, p$ reaction
at $E_\gamma^{Lab}=\,$ 4 GeV  for a total center of mass energy of the $\eta p$ pair of 1.54 GeV.
\begin{figure}[h]
\centerline{\psfig{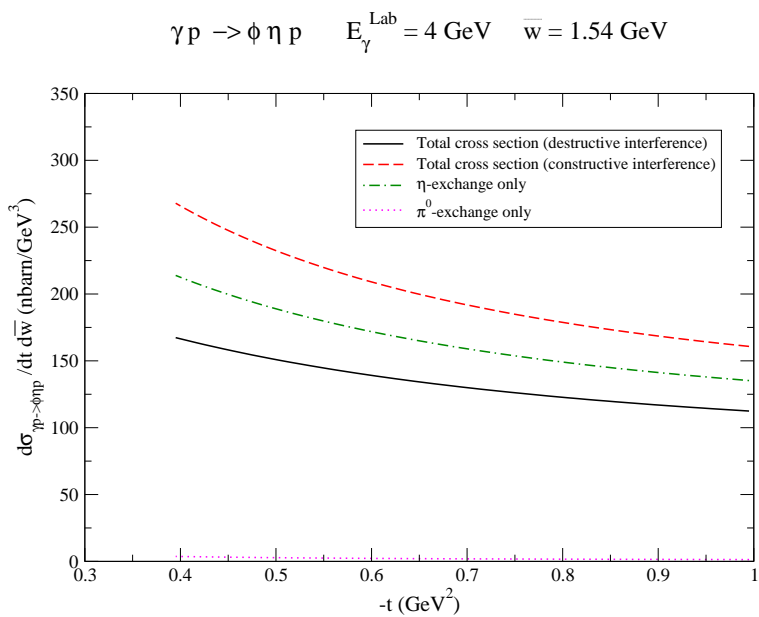}}
\vspace*{8pt}
\caption{Differential cross section
${d\sigma_{\gamma \,p\rightarrow \phi \, \eta \,p}}/ {dt \,d{\bar w}}$
at $E_\gamma^{Lab}=\,$ 4 GeV for a total center of mass energy of the $\eta p$ pair of 1.54 GeV.
The full and dashed lines are the total differential cross sections assuming a destructive and a constructive
$\pi^0$-$\eta$ interference respectively. The dot-dashed and dotted lines show the contributions of the
$\eta$-meson and $\pi^0$-exchanges (from Ref. 4).}
 \label{f5}
\end{figure}

The features of the results are quite similar to those observed on Fig. 4. The main difference is that
the cross section is roughly twice larger, as a consequence of the opening of phase space for the $\eta N$ final state.

The main conclusions to be drawn from this calculation can be summarized as follows.

In the
kinematics where the $\eta p$ invariant mass in the final state
lies between the threshold value (m$_p$+m$_\eta$) and the N*(1535) resonance mass,
the
$\gamma \, p \rightarrow \phi \, \eta \, p$ reaction is largely determined by the $\eta N$
scattering amplitude close to threshold. The most appropriate range of initial photon
energy is  $4\, <\,E_\gamma^{Lab}\,<\,5$ GeV,
in order to reach low (absolute) values of the squared 4-momentum transfer and to
be able to rely on meson-exchange pictures.

The actual magnitude of the $\gamma \, p \rightarrow \phi \, \eta \, p$ reaction
cross section depends significantly on the relative sign
of the coupling constants
$g_{\phi\pi\gamma}$ and $g_{\phi\eta\gamma}$. An independent determination of that sign would
be very useful.

Finally, accurate data on the $\gamma \, p \rightarrow \phi \, \eta \, p$ reaction
taken at  $4\, <\,E_\gamma^{Lab}\,<\,5$ GeV and involving t-distributions at different
$\eta p$ invariant masses would clearly contribute to the understanding of the
$\eta p$ scattering amplitude in the threshold region.

\vskip 0.5 true cm
\noindent{\bf Acknowledgement}\par
\medskip
\noindent
One of us (M. S.) thanks the organizers of the Meson2006 Workshop for giving her
the opportunity to present this work in a stimulating context where $\eta$-nucleon physics close to threshold
was thoroughly discussed.
We acknowledge the support of the European Community-Research Infrastructure Activity under
the FP6 "Structuring the European Research Area" programme
(Hadron Physics, contract number RII3-CT-2004-506078).


\begin{thebibliography}{0}    
\bibitem{Lutz1} M.F.M. Lutz, Gy. Wolf, B. Friman, {\it Nucl. Phys. A} {\bf 706}, 431 (2002), ERRATUM-ibid
{\it A} {\bf 765}, 495 (2006).
\bibitem{GreenWycech}
A.M. Green, S. Wycech, {\it Phys. Rev. C} {\bf 71}, 014001 (2005).
\bibitem{PDG}
S. Eidelman et al., {\it Phys. Lett. B} {\bf 592}, 1 (2004).
\bibitem{Lutz2} M.F.M. Lutz, M. Soyeur, {\it Nucl. Phys. A} {\bf 773} 239.
\bibitem{Barth}
J. Barth et al., {\it Eur. Phys. J. A} {\bf 17}, 269 (2003).
\bibitem{CLAS}
K. McCormick et al., {\it Phys. Rev. C} {\bf 69}, 032203 (2004).
\bibitem{Mibe}
T. Mibe et al., {\it Phys. Rev. Lett.} {\bf 95}, 182001 (2005).
\bibitem{Titov}
A.I. Titov and T.-S.H. Lee, {\it Phys. Rev. C} {\bf 67}, 065205 (2003).



\end{thebibliography}
\end{document}